# Short linear motif candidates in the cell entry system used by SARS-CoV-2 and their potential therapeutic implications

Bálint Mészáros[1] (ORCID: 0000-0003-0919-4449)
Hugo Sámano-Sánchez[1] (ORCID: 0000-0003-4744-4787)
Jesús Alvarado-Valverde[1,2] (ORCID: 0000-0003-0752-1151)
Jelena Čalyševa[1,2] (ORCID: 0000-0003-1047-4157)
Elizabeth Martínez-Pérez[1,3] (ORCID: 0000-0003-4411-976X)
Renato Alves[1] (ORCID: 0000-0002-7212-0234)
Manjeet Kumar[1] (ORCID: 0000-0002-3004-2151) Email: manjeet.kumar@embl.de
Friedrich Rippmann[4] (ORCID: 0000-0002-4604-9251)
Lucía B. Chemes[5] (ORCID: 0000-0003-0192-9906) Email: lchemes@iib.unsam.edu.ar
Toby J. Gibson[1] (ORCID: 0000-0003-0657-5166) Email: toby.gibson@embl.de

[1]Structural and Computational Biology Unit, European Molecular Biology Laboratory, Heidelberg 69117, Germany
[2]Collaboration for joint PhD degree between EMBL and Heidelberg University, Faculty of Biosciences
[3]Laboratorio de bioinformática estructural, Fundación Instituto Leloir, C1405BWE, Buenos Aires, Argentina
[4]Computational Chemistry & Biology, Merck KGaA, Frankfurter Str. 250, 64293 Darmstadt, Germany
[5]Instituto de Investigaciones Biotecnológicas, Universidad Nacional de San Martín. Av. 25 de Mayo y Francia S/N, CP1650 Buenos Aires, Argentina

## Abstract

The primary cell surface receptor for SARS-CoV-2 is the angiotensin-converting enzyme 2 (ACE2). Recently it has been noticed that the viral Spike protein has an RGD motif, suggesting that cell surface integrins may be co-receptors. We examined the sequences of ACE2 and integrins with the Eukaryotic Linear Motif resource, ELM, and were presented with candidate short linear motifs (SLiMs) in their short, unstructured, cytosolic tails with potential roles in endocytosis, membrane dynamics, autophagy, cytoskeleton and cell signalling. These SLiM candidates are highly conserved in vertebrates. They suggest potential interactions with the AP2 µ2 subunit as well as I-BAR, LC3, PDZ, PTB and SH2 domains found in signalling and regulatory proteins present in epithelial lung cells. Several motifs overlap in the tail sequences, suggesting that they may act as molecular switches, often involving tyrosine phosphorylation status. Candidate LIR motifs are present in the tails of ACE2 and integrin β3, suggesting that these proteins can directly recruit autophagy components. We also noticed that the extracellular part of ACE2 has a conserved MIDAS structural motif, which are commonly used by β integrins for ligand binding, potentially supporting the proposal that integrins and ACE2 share common ligands. The findings presented here identify several molecular links and testable hypotheses that might help uncover the mechanisms of SARS-CoV-2 attachment, entry and replication, and strengthen the possibility that it might be possible to develop host-directed therapies to dampen the efficiency of viral entry and hamper disease progression. The strong sequence conservation means that these putative SLiMs are good candidates: Nevertheless, SLiMs must always be validated by experimentation before they can be stated to be functional.

## Introduction

The COVID-19 pandemic is caused by severe acute respiratory syndrome coronavirus 2 (SARS-CoV-2), an enveloped, single-stranded RNA virus. It had infected more than two and a half million people and caused circa 170,000 deaths globally by mid-April 2020. SARS-CoV-2 belongs to the Coronaviridae family, whose members are common human pathogens responsible for the common cold, as well as for some emerging severe respiratory diseases. Among them are the SARS-CoV and the Middle East respiratory syndrome coronavirus (MERS-CoV), the former of which caused over 8,000 cases in 2003 with a fatality rate of ~10%, and the latter caused about 2,500 infections in 2012 with a fatality rate of 37% (de Wit et al., 2016). Another coronavirus, infectious bronchitis virus (IBV), infects birds and has been used as a model in coronavirus research (Sisk et al., 2018). SARS-CoV-2, like SARS-CoV (Li et al., 2003), uses the angiotensin converting enzyme 2 (ACE2) as a receptor to attach to the host cells (Hoffmann et al., 2020; Wrapp et al., 2020; P. Zhou et al., 2020). ACE2 is a single pass type I membrane protein with a short cytosolic C-terminal region for which the functionality, however, is mostly unknown.

Earlier results clearly show that the SARS-CoV-2 receptor binding domain (RBD) of the Spike protein interacts with ACE2 for cellular entry. However, type II alveolar cells (AT2) – the main targets of SARS-CoV-2 in the lung (Zou et al., 2020) – express a relatively low amount of ACE2, which points to the existence of co-receptors being targeted by the virus in parallel. One such candidate are integrins that bind a large variety of ligands harbouring an RGD sequence motif, as recent analysis of the RBD identified a possibly functional RGD motif (Sigrist et al., 2020).

Integrins are major cell attachment receptors, which are known to be targeted by a range of viruses - including HIV, herpes simplex virus-2, Epstein-Barr virus (EBV) and the foot and mouth virus - for cell entry and activation of linked intracellular pathways (Hussein et al., 2015; Stewart and Nemerow, 2007; Triantafilou et al., 2001). Integrins are special types of receptors, as they propagate signals in both directions; extracellular ligands can induce cytoplasmic pathway activation, but intracellular binding on the cytosolic tails can influence the structure of the ectodomains and hence ligand-binding affinity. The complexity of integrin signalling stems in the dimeric structure of integrins, as they are composed of two subunits, α and β. For the RGD-binding integrins, the ligand binding surface lies at the interface of the two integrin subunits with both subunits making contacts with the ligand. These RGD motifs are recognized by at least 8 out of the 24 human integrins, and the flanking residues next to the core RGD motif are known to play a huge role in selectivity (Kapp et al., 2017). Several viral proteins contain RGD (or RGD-like) short linear motifs (SLiMs) for integrin modulation; and in addition, some viruses can not only use integrins on the host cell surface, but HIV and SIV can also incorporate integrins into their own membranes for mediating interactions with the host (Guzzo et al., 2017). Therefore, integrins can potentially be targeted at both the extracellular and the intracellular side to combat pathogenic hijacking.

Viruses, as obligate intracellular entities, need to interfere with major cellular processes like vesicular trafficking, cell cycle, cellular transport, protein degradation or signal transduction to satisfy their replication, enzymatic, metabolic and transport needs (Davey et al., 2011). To achieve this, a large number of host processes are hijacked using SLiMs often located in intrinsically disordered regions to establish protein-protein interactions with host proteins or undergo post-translational modifications, like tyrosine phosphorylation. For example, cellular signalling relies heavily on the use of SLiMs (Van Roey et al., 2013, 2012). The low affinity and cooperativity of SLiM-based molecular processes allow reversible and transient interactions that can work as switches between distinct functional states and are regulated both in time and space (Gibson, 2009; Scott and Pawson, 2009). Conditional switching of SLiMs, for example through phosphorylation, can induce the exchange of binding partners for a protein, thus mediating molecular decision-making in response to signals reporting on the cell state (Van Roey et al., 2012). The central resource for SLiMs is the Eukaryotic Linear Motif database (ELM; http://elm.eu.org/), also serving as an exploratory server for over 280 manually curated SLiM classes with experimental evidence, each defined by a POSIX regular expression (Kumar et al., 2020).

As explained above, a major strategy of viruses is to abuse the host system by using mimics of eukaryotic SLiMs to compete with extracellular or intracellular binding partners or to sequester host proteins (Davey et al., 2011). This dependence of viruses and many other pathogens on SLiM-mediated functions suggests that there is an opportunity to drug the cell systems where these interactions are being hijacked (Sámano-Sánchez and Gibson, 2020). For example, tyrosine kinase inhibitors, often used in anticancer therapy, have shown promising coronavirus replication inhibition in infectious cell culture systems (Coleman et al., 2016; Dong et al., 2020; Shin et al., 2018; Sisk et al., 2018). In the remainder of the introduction, we will describe some of the major pathways hijacked by viruses to accomplish cell attachment, entry and replication, which are suggested by our results to be relevant to SARS-CoV-2 infection.

Receptor Mediated Endocytosis (RME) is a cellular import process triggered by cell surface receptor proteins, including any cargoes attached to them, in which a large vesicular structure is assembled entirely through cooperative low affinity interactions of SLiMs and phospholipid head groups with their globular protein domain partners. The vesicles are strong and stable, yet flexible and dynamically assembled and disassembled. The external triggering of surface receptors (many of which have the YxxPhi or NPxY tyrosine sorting motifs) is transmitted across the plasma membrane, inducing local enzymatic modification of lipid head-groups from PI4P to (PI(4,5)P2) by the PIPK1 kinase. The local enrichment of (PI(4,5)P2) enables binding of domains such as ENTH in Epsins which can begin to curve the membrane and assemble clathrin cages using their Clathrin Box motif and also attract additional adapter proteins via yet more SLiMs. In turn, additional sets of SLiM-bearing proteins stimulate the actin filament formation and attachment, necessary to fold and pull the invagination into the cytosol. Later, Dynamin binds directly to (PI(4,5)P2) on the membrane to complete the scission process. Once in the cytosol, the clathrin-coated vesicles are soon dismantled and the contents are included into the early endosomes. (For recent reviews of the process, see (Haucke and Kozlov, 2018; Kaksonen and Roux, 2018; Senju and Lappalainen, 2019). Many viruses enter the cell via endocytosis, using many different cell surface receptors (Grove and Marsh, 2011). Viruses such as HIV and Hepatitis C virus depend on the recognition of more than one receptor for entry, but in many cases the stoichiometry of receptor engagement is unknown. Coronaviruses can enter cells through different routes that include receptor mediated endocytosis or fusion at the membrane (de Haan and Rottier, 2006). In the case of SARS-CoV, the main entry route is endocytic, and depends on endosome acidification (Huang et al., 2006; Simmons et al., 2004). However, protease-mediated activation of the Spike protein relieves the pH-dependence of viral entry, indicating that acidification is not a requirement per se, but acts by inducing the endosomal Spike protein cleavage required for viral fusion (Matsuyama et al., 2005; Simmons et al., 2005). Spike protein cleavage can be done by the transmembrane protease serine 2 (TMPRSS2) at the cell surface, or by Cathepsin L within endosomes (de Wilde et al., 2018). The same entry route and proteases are utilized by SARS-CoV-2, and the main entry route also seems to be endocytic (Hoffmann et al., 2020; Ou et al., 2020).

Autophagy is an evolutionarily conserved process in eukaryotes with multiple cellular roles that include the regulation of cellular homeostasis through the catabolism of cell components, immune development and the host cell response to infection through pathogen phagocytosis (Deretic and Levine, 2009). Viruses have evolved mechanisms to block the host cell antiviral response, and can further hijack autophagy components to promote their survival and replication. This can be done through viral mimicry of host proteins coordinating autophagy or through the direct inhibition of the host autophagy machinery (Kudchodkar and Levine, 2009). Coronaviruses (CoVs) exploit the autophagy machinery through different mechanisms (Cong et al., 2017; Maier and Britton, 2012). For example, MERS-CoV targets the BECN1 autophagy regulator for degradation, blocking the fusion of autophagosomes and lysosomes and protecting the virus from degradation (Gassen et al., 2019). Coronaviruses repurpose cellular membranes to create double membrane vesicles (DMVs) onto which the replication-transcription complex (RTC) is assembled, a process that involves recruitment of multiple autophagy components (Cong et al., 2017; Prentice et al., 2004; V'kovski et al., 2019). Betacoronavirus mouse hepatitis virus (MHV) RTCs assemble by recruiting LC3-I, a non-lipidated form of the LC3 autophagy protein (Cong et al., 2017; Reggiori et al., 2010), and SARS-CoV RTCs also colocalize with LC3 (Prentice et al., 2004). Proximity-based mass spectrometry on the MHV replication complex further revealed that the RTC environment repurposes components from the host autophagy, vesicular trafficking and translation machineries (V'kovski et al., 2019)

In the present work, we identify a set of conserved SLiM candidates in the ACE2 and integrin proteins, which are likely to act in the cell entry system of SARS-CoV-2. These motifs can provide molecular links to understand how the virus recognizes target membranes, enters into cells, and how it repurposes intracellular membrane components to drive its replication. These molecular links might provide novel clues towards drugging SARS-CoV-2 infections. We first focus on the extracellular SLiMs, before moving across the membrane to examine the cytosolic potential of the receptor tails.

# Extracellular receptor interplay and viral hijacking in the ACE2/integrin system

### *Integrins are candidates for acting as co-receptors for SARS-CoV-2 entry*

The ability of SARS-CoV-2 RBD to bind integrins via the RGD motif (see Table 1) has not yet been assessed directly. However, there are several features that make the Spike-integrin interaction plausible, including sequence- and structure-level information, expression profiles, the presence of accessory motifs and protein-protein interactions.

### *Evolution of the RGD motif in Spike receptor binding domain*

Aligning close homologues of the RBD from the Coronaviridae family (Figure 1A) shows that the motif candidate is located in a locally less conserved region, hinting at the rapid evolvability of the site. Several coronavirus RBDs contain KGD at this site, which is known to be a lower affinity integrin binding site, first identified in snake venom disintegrins, such as barbourin (Minoux et al., 2000). A KGD motif residing in the EBV gH/gL protein has also been shown to be essential for entry into epithelial and B cells (Chen et al., 2012). Figure 1B shows the tree derived from the Spike protein sequence alignment, highlighting that the SARS-CoV-2 RGD motif might have evolved from an earlier KGD motif, and might present a distinct step of adaptation if the motif is indeed an integrin attachment site.

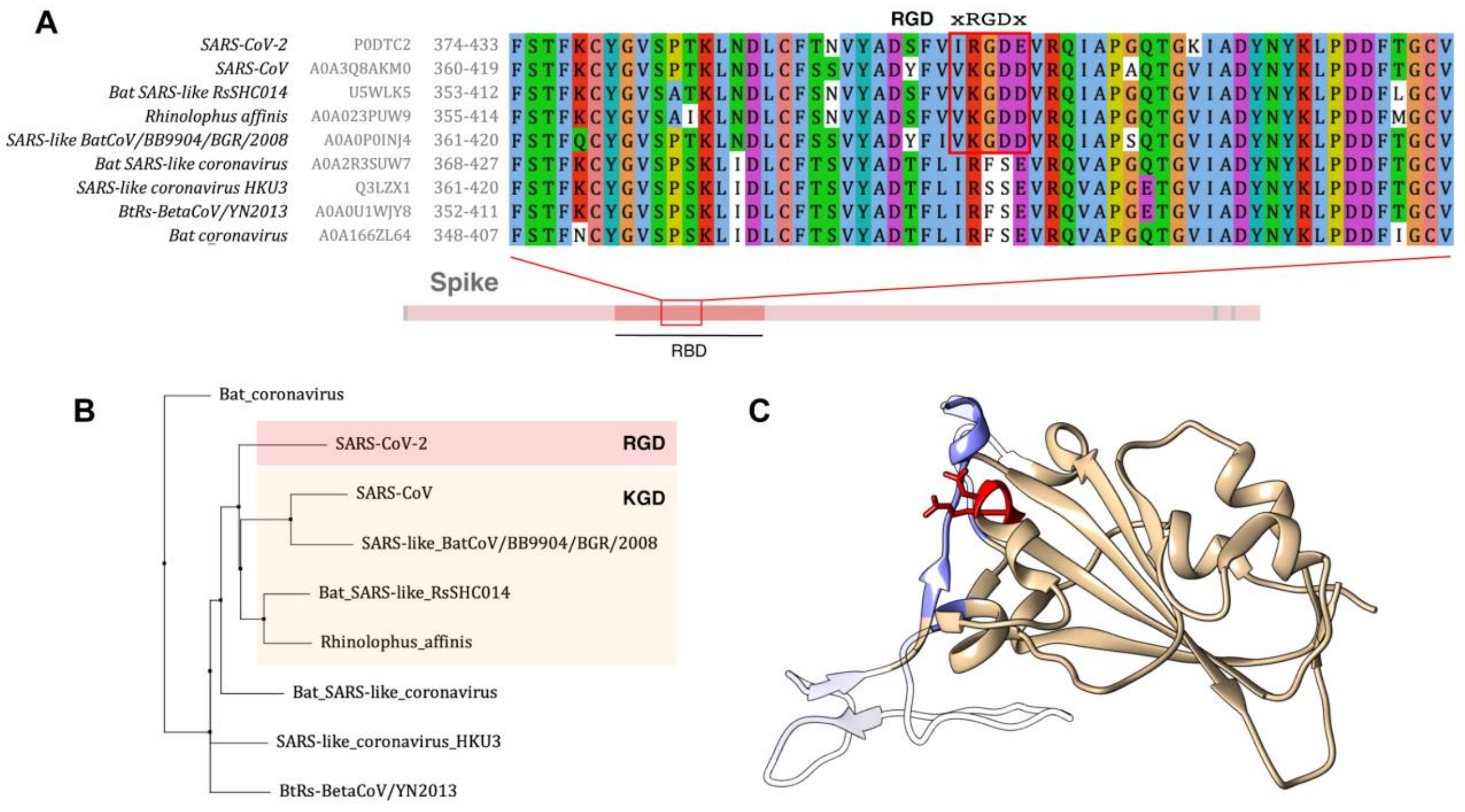

***Figure 1: The RGD motif of the SARS-CoV-2 spike protein.*** *A) Multiple sequence alignment of a part of the SARS-CoV-2 Spike RBD region using 9 homologous sequences from closely related viruses and showing the conservation of the RGD motif. Red box marks the conservation range of the [RK]GD motif. Viruses' names, UniProt IDs and sequence numberings are listed on the left side of the alignment. The location of the region shown in the alignment is indicated in a representative diagram of the Spike protein. B) Guide neighbour joining tree of the multiple sequence alignment with the sequences containing the potential low- and high-affinity integrin binding motifs (KGD and RGD) shown in orange and red boxes, respectively. C) Structure of the SARS-CoV-2 RBD as seen in the ACE2-bound form (PDB ID:6m17). The RGD motif is shown in red sticks. Regions in direct contact with ACE2 are shown in blue. Residues with missing atomic coordinates (indicating flexibility) in the unbound trimeric Spike protein structures (PDB IDs: 6vsb, 6vxx and 6vyb) are shown in transparency. Alignment and tree prepared in Jalview (Waterhouse et al., 2009) with Clustal colours. Structure was visualized using UCSF Chimera (Pettersen et al., 2004).*

### *The RBD-integrin interaction is structurally feasible*

At the time of reporting the RGD motif, no SARS-CoV-2 Spike structures were available, so the authors used structural homology modelling to determine that the RGD motif is surface accessible (Sigrist et al., 2020). Since then, several RBD structures have been determined, both in unbound (Walls et al., 2020; Wrapp et al., 2020) and ACE2 complexed forms using electron microscopy (Yan et al., 2020) and X-ray diffraction (Lan et al., 2020), allowing for the direct structural assessment of the possibility of binding to integrins. Figure 1C shows the RGD motif together with the residues involved in direct binding of ACE2. The RGD motif is located in the vicinity of the ACE2 binding site, however, based on uncomplexed structures of the RBD, the residues that surround the RGD site are flexible. This indicates that while ACE2 binding blocks the RGD motif, without ACE2 the RGD is surface accessible and interaction with integrins are not sterically blocked.

As Spike exists as a trimer on the virion surface, different copies of the RBD can in theory interact with ACE2 and integrins at the same time. There is no solved complex structure of the ACE2-integrin complex. However, further structural consideration may indicate whether the Spike-ACE2 and the Spike-integrin interactions can coexist. The ectodomains of both ACE2 and integrins in the open conformation are roughly the same size measured from the membrane, being in the 150-200 Å range (based on available structures PDB:6m17 (Yan et al., 2020) and PDB:3ije (Xiong et al., 2009)). This means that the RGD-binding site of integrins and the RBD binding regions of ACE2 are relatively close in space. In addition, in the more open ‘up’ conformation of the RBDs, the distance between pairs of RBDs is about 100 Å (based on the structure PDB:6vsb reported in (Wrapp et al., 2020)), which is probably larger than the distance between the ACE2 binding region and the integrin ligand binding site, estimated from the individual integrin and ACE2 structures.

### *α5β1 and αvβ3 integrins are potential targets for SARS-CoV-2 RBD*

The RGD motif is recognized by several integrins, and specificity is determined mostly by the flanking residues around the core motif. As evidenced by crystallized integrin dimer:ligand complexes, the residue preceding RGD is in contact with the α subunit, while the residue after the core motif interacts with the β subunit. The immediate context of the SARS-CoV-2 RBD motif is 402-I**RGD**E-406, which can give an indication about possible integrin targets. I**RGD** can be found in several native integrin-binding partners, including FREM1 (Kiyozumi et al., 2012), MFAP4 (Pilecki et al., 2015) and IGFBP1/2 (Cavaillé et al., 2006; Wang et al., 2006). These extracellular matrix proteins target integrins with αv, α5 and α8 subunits. **RGD**E is present in the native human integrin ligands TGF-β1, osteolectin, collagen α-1(VI) chain, PSBG-9 and polydom, and *in vitro* and *in vivo* binding studies on the specificity profiles of these proteins (Cescon et al., 2015; Rattila et al., 2019; Sato-Nishiuchi et al., 2012; Shanley et al., 2013; Shen et al., 2019; Tumbarello et al., 2012) highlighted a post-RGD Glu to be efficient in binding to β1, β2 and β3 integrin subunits. Correlating these preferences with possible α and β integrin subunit pairings points to the most likely candidate target integrins for SARS-CoV-2 being αvβ1, αvβ3, α5β1 and α8β1.

Motif-domain interactions are typically under heavy spatio-temporal regulation. Hence the SARS-CoV-2 RBD-integrin binding can only occur if the possible target integrins are expressed on AT2 cells. Both α5β1 (Pilewski et al., 1997) and αvβ3 (Caccavari et al., 2010; Nakamura et al., 2002; Singh et al., 2000) have been observed in lung epithelial cells and have been shown to be implicated in disease emergence and progression, including emphysema, non-small cell lung cancer and mechanical injury of the lungs (Teoh et al., 2015), marking these two integrins as prime suspects for targets of the RBD.

### *Native ACE2/integrin interplay*

It has been shown that in heart tissues, ACE2 is able to bind the β1 subunit of integrins in an RGD-independent manner, enhancing cell adhesion and regulating integrin signalling via the focal adhesion kinase (FAK) (Clarke et al., 2012). The RGD independence of the interaction means that while ACE2 and integrins are in complex, the RGD binding site of the integrin is unoccupied, further supporting a potential integrin:ACE2:Spike ternary interaction.

Apart from the known interplay between ACE2 and integrins, there are additional features that indicate an even tighter crosstalk between the two receptors. RGD-mediated binding to integrins is metal mediated (via divalent cations like $Mg^{2+}$ or $Mn^{2+}$), and all integrins have a so-called ‘metal ion-dependent adhesion site’ (MIDAS) motif (DxSxS) (Lee et al., 1995). The integrin MIDAS structural motif is located near the ligand binding site on the β subunit and is essential for binding, as sidechains belonging to the motif and an acidic residue from the ligand coordinate the metal ion together (Zhang and Chen, 2012). ACE2 also has a similar DxSxS motif (see Table 1), that might facilitate interactions with ligands that are recognized by integrins, possibly creating an overlap between the ligand binding profiles and regulation of the two receptors. In the known structures where Spike is bound to ACE2 the RGD motif is not in contact with the ACE2 MIDAS (Yan et al., 2020). However, the MIDAS motif is highly conserved across species (see Figure 2), and surface exposed. Consequently it may still be involved in mediating an interaction with an RGD-like motif, potentially serving as a parallel mechanism for binding the Spike protein.

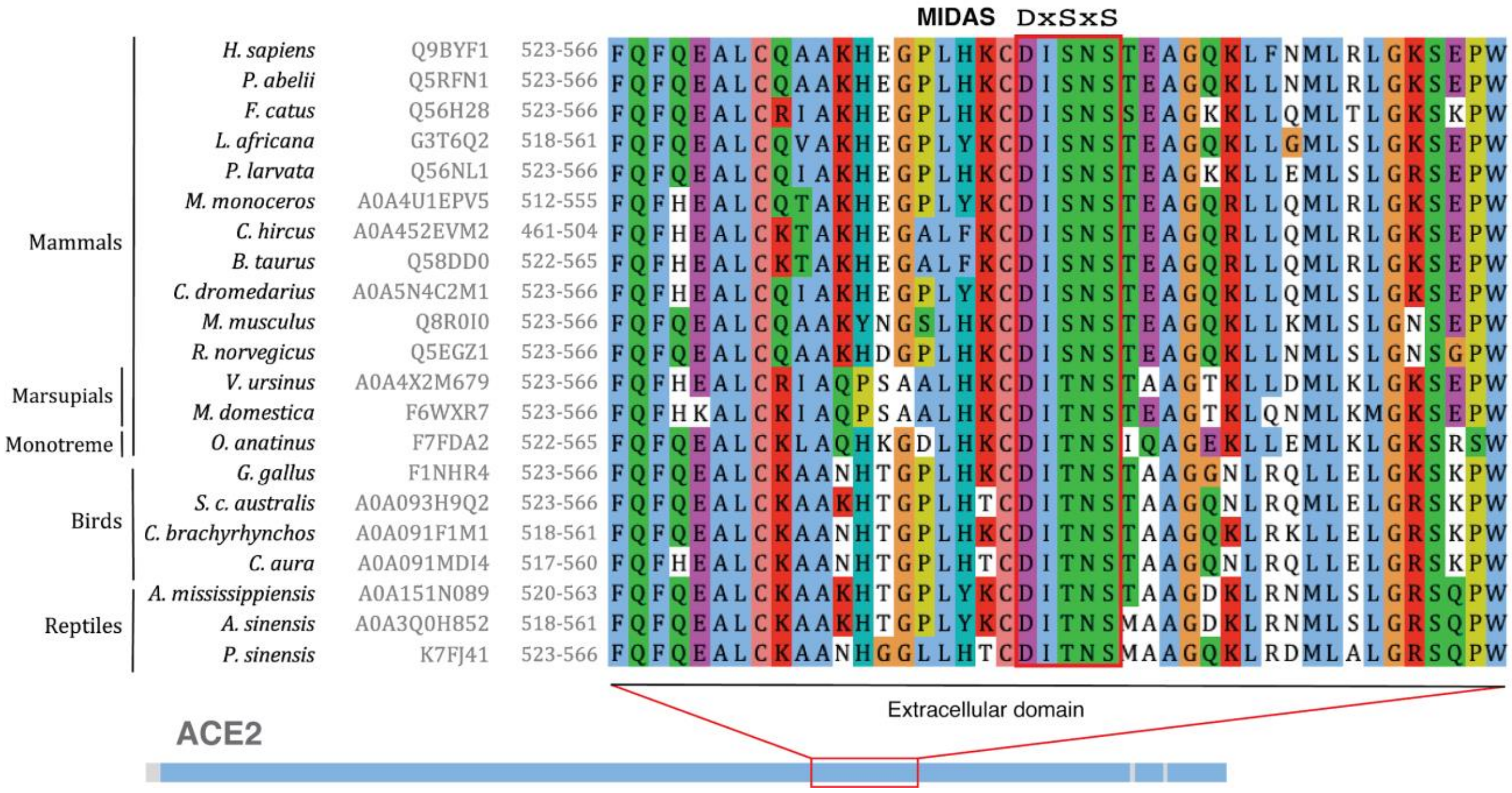

***Figure 2: Alignment of ACE2 illustrating conservation of the MIDAS motif****. Multiple sequence alignment of a part of the ACE2 extracellular region using 21 homologous sequences from different vertebrates (mammals, birds and reptiles) and showing the conservation of the Dx[ST]xS motif (main residues displayed above). Red boxes mark the conservation range of the MIDAS motif in all sequences. Organism names, UniProt IDs and sequence numberings are listed on the left side of the alignment. The location of the region shown in the alignment is indicated in a representative diagram of the ACE2 protein. Figure prepared with Jalview using Clustal colours.*

### *Protease usage by SARS-CoV-2*

ACE2 and several integrin subunits require proteolytic cleavage for biological activity (see Table 1). Integrin subunits α3, α5, α6 and αv are cleaved by furin or furin-like proprotein convertases (PCs) during maturation (Lehmann et al., 1996; Lissitzky et al., 2000). Nearly all PCs contain an RGD motif, and while its role in integrin binding is not clear, the motif has been shown to be required for proper functioning for several PCs (Lou et al., 2007; Lusson et al., 1997; Rovère et al., 1999). The SARS-CoV-2 Spike protein contains a furin-like cleavage site that is absent from closely related Spike proteins, immediately following the RBD (Coutard et al., 2020). This cleavage results in increased virulence, possibly by allowing greater movement of the RBD potentially aiding in exploring a larger space around the RBD-binding region of ACE2.

ACE2 is cleaved by several proteases, including TMPRSS2 (Heurich et al., 2014). ACE2 binds to TMPRSS2, forming a receptor-protease complex (Shulla et al., 2011). TMPRSS2 is also known to cleave the Spike protein of both SARS-CoV and MERS-CoV (Iwata-Yoshikawa et al., 2019), augmenting their entry into the host cell (Heurich et al., 2014). Furthermore, similar results have been found for SARS-CoV-2, where TMPRSS2 was found to be fundamental for cell entry (Hoffmann et al., 2020). This dependence is most probably two-fold: on one hand TMPRSS2 is needed for ACE2 activation, on the other hand, SARS-CoV-2 Spike protein also contains a TMPRSS2 cleavage site (Meng et al., 2020).

## SLiM candidates in the ACE2 receptor intrinsically disordered tail

The ACE2 sequence (UniProt:ACE2_HUMAN) was entered in the ELM server (Kumar et al., 2020) and returned several relevant candidate SLiMs in the short cytosolic C-terminal tail. Because SLiMs are so short, it is difficult to obtain significant results in sequence searches. Contextual information, including cell compartment localisation and functional relevance, is important in deciding whether a motif candidate is worth testing experimentally (Gibson et al., 2015). Furthermore, in intrinsically unstructured protein sequence, amino acid conservation is indicative of functional interactions. Therefore, an alignment was prepared of vertebrate ACE2 proteins. All of the detected motif matches (shown in Table 1 together with potential binding partner domains defined using Pfam (El-Gebali et al., 2019) and InterPro (Mitchell et al., 2019)) were conserved in mammals, most were conserved with birds and mammals and some were conserved with extant reptiles (Figure 3). These groups diverged from one another >300 million years ago (Kemp, 2005) indicating a strong conservation of all candidate motifs. In addition, the functional contexts of these motifs are biologically coherent, involving signalling by tyrosine kinases, endocytosis, autophagy and actin filament induction (Table 1). In the following subsections we briefly summarise each of the conserved motifs and their possible role in the viral entry mechanism.

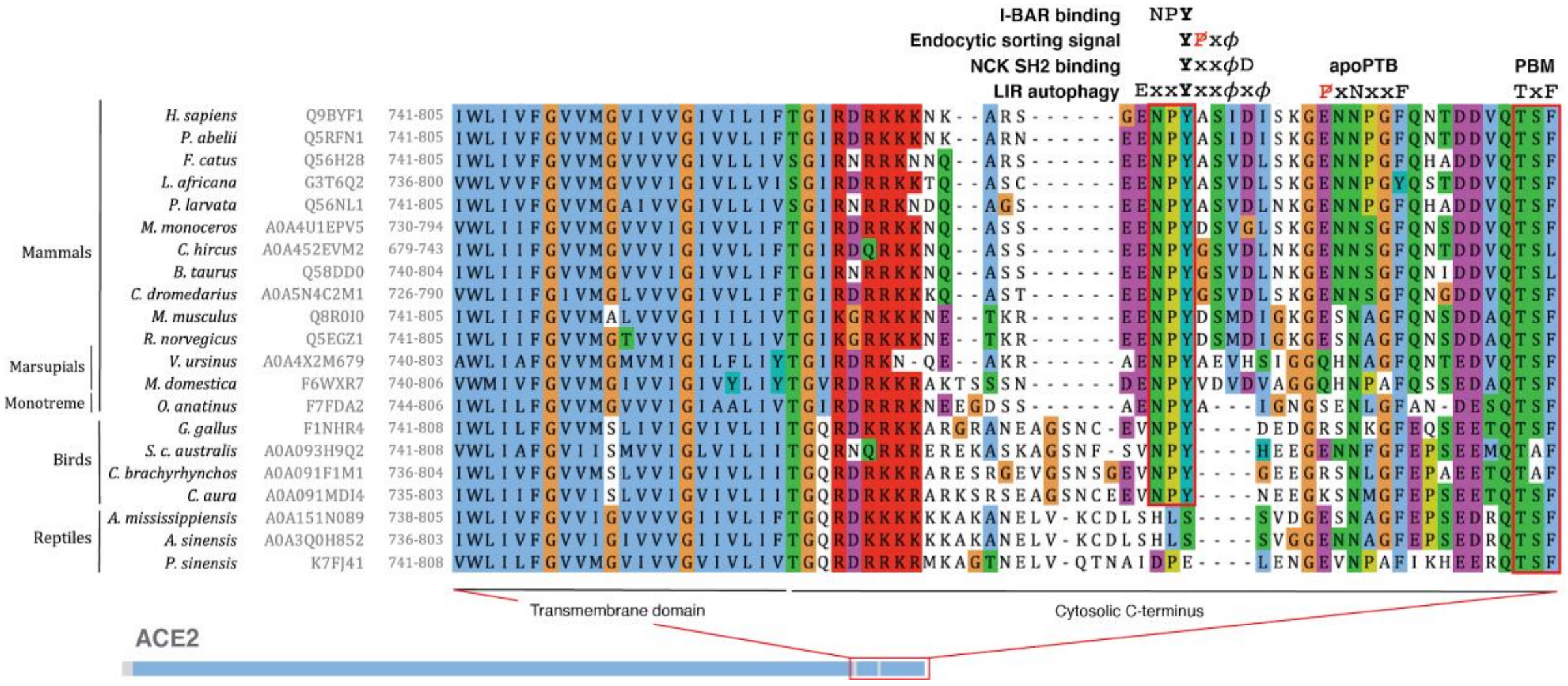

***Figure 3: Alignment of ACE2 illustrating conserved motifs in the cytosolic C-terminal tail following the transmembrane helix**. Multiple sequence alignment of ACE2 transmembrane and C-terminus regions using 21 homologous sequences from different vertebrates (mammals, birds and reptiles) and showing their motif conservation. The names (bold) and key residues of the motifs are displayed above the alignment (ϕ stands for a bulky hydrophobic residue), including a conserved tyrosine (bold) and excluded positions (red and crossed). Red boxes mark the conservation range of the PDZ-binding motif (PBM) (all sequences) and NPY motif (in mammals and birds). Organism names, UniProt IDs and sequence numberings are listed on the left side of the alignment. The location of the region shown in the alignment is indicated in a representative diagram of the ACE2 protein. Figure prepared with Jalview using Clustal colours.*

### *The YxxPhi endocytic sorting signal*

The YxxPhi motif binds the μ2 subunit (UniProt:AP2M1_HUMAN) of the endocytosis AP-2 adaptors by β-augmentation (Owen and Evans, 1998). It is found in numerous cell surface receptors which have intrinsically disordered C-terminal tails (Bonifacino and Traub, 2003). A small selection is listed in the database entry ELM:TRG_ENDOCYTIC_2, and while the motif has not been validated in ACE2, it is highly conserved (Figure 3). When the Tyr is phosphorylated, this motif becomes an SH2 binding site, while in the apo form it binds the μ2 adapter. Therefore, this motif can operate as a molecular switch. The residue following the Tyr makes a β-strand interaction and therefore cannot be a proline (PDB:1bxx). The Phi position requires a bulky hydrophobic residue. The motif pattern can be represented by the regular expression Y[^P].[LMVIF] and this motif is conserved in ACE2 from all mammals except monotremes. Thus the mammalian ACE2, which internalises the coronavirus, has a SLiM candidate for internalisation appropriately located within its cytosolic tail.

### *The NCK SH2 binding motif*

The region encompassing the YxxPhi motif overlaps with a Src homology 2 (SH2) domain binding motif (Figure 3) that is created upon phosphorylation of Tyr 781. SH2 domain binding motifs are characterized by an invariant phosphotyrosine (pY) that is created following tyrosine kinase activation, and allows binding to more than 100 types of SH2 domains present in human proteins (Tinti et al., 2013). The pY residue is accompanied by additional binding determinants that frequently involve hydrophobic residues at the pY+3 position, but can also involve other combinations, such as Asn at pY+2 in Grb2-specific SH2 motifs, or hydrophobic residues at pY+4 in STAP-1 SH2 motifs (Huang et al., 2008; Kaneko et al., 2010). Most motifs are also characterized by the exclusion of residues at certain positions following the pY, and in general, SH2-binding motifs show a high degree of cross-specificity (Huang et al., 2008; Liu et al., 2010).

Cell culture infection assays with different Coronaviruses, including SARS-CoV, have shown susceptibility to tyrosine kinase inhibitors, indicating the involvement of host tyrosine phosphorylation (Coleman et al., 2016; Dong et al., 2020; Shin et al., 2018; Sisk et al., 2018). The sequence found in ACE2 (781-YASID-785) best matches the binding specificity for the SH2 domain present in NCK1/2 proteins, which belong to the Class IA SH2 domains (Kaneko et al., 2010). Proteins known to contain this motif are listed in entry ELM:LIG_SH2_NCK1_1. NCK proteins are adaptor proteins that modulate actin cytoskeleton dynamics (Buday et al., 2002). For example, NCK is recruited to the cell membrane by the Nephrin protein, following tyrosine signalling that creates several NCK SH2 binding motifs in Nephrin (Blasutig et al., 2008). Once recruited to the membrane, NCK activates Wiskott-Aldrich syndrome (WASP) family proteins through the use of a helical binding motif (ELM:LIG_GBD_CHELIX_1) that relieves WASP autoinhibition and allows the recruitment of the actin regulatory protein complex Arp2/3, which leads to the initiation of actin polymerization (Okrut et al., 2015). The NCK binding motif is exploited by two known human pathogens, namely enteropathogenic *Escherichia coli* and vaccinia virus (Frese et al., 2006). The residues present at pY+1, pY+2 and pY+4 rule out that the ACE2 YASID motif can be a strong Grb2, CRK or STAP1 SH2 domain binder, and binding to STAT1/2/3 SH2 domains is also unlikely due to the lack of adequate specificity determinants. Other SH2 domains (*e.g.* Src-related) could be recruited by ACE2, and experimental validation will be required to test these hypotheses.

### *The NPY IBAR binding motif*

Tyrosine 781 in ACE2 also overlaps a phosphorylation-independent NPY motif (ELM:LIG_IBAR_NPY_1). This motif was initially described in the bacterial secreted protein translocated intimin receptor (Tir) from pathogenic strains of *E. coli* like enterohaemorrhagic *E. coli* (EHEC). The NPY tripeptide recognizes and binds with a 60 μM affinity to inverse Bin-Amphiphysin-Rvs (I-BAR) domains in adaptor proteins like insulin receptor substrate protein of 53 kDa (IRSp53) and its homolog insulin receptor tyrosine kinase substrate (IRTKS) (Campellone et al., 2006; de Groot et al., 2011). I-BAR domains bind to the plasma membrane to favour weak membrane protrusions, and the preference of I-BAR domains for negative membrane curvatures enables a positive feedback loop that can result in the formation of lamellipodia, filopodia and other types of membrane protrusions (Chen et al., 2015; Prévost et al., 2015; Zhao et al., 2011). IRSp53 and IRTKS are modular proteins that contain SH3 domains which in turn recognize PxxP SLiMs in actin filament regulators like Mena, Eps8 and mDia1 (Ahmed et al., 2010) resulting in the formation of membrane protrusions through actin filament formation (Campellone et al., 2006; Chen et al., 2015; Prévost et al., 2015; Zhao et al., 2011). Moreover, IRSp53 has an additional Cdc42-binding motif that can result in a direct WASP activation (Ahmed et al., 2010). During EHEC infection, the bacteria uses the NPY motif in the transmembrane protein Tir to recruit IRSp53 (Campellone et al., 2006). IRSp53 acts as a scaffold to localize the injected bacterial protein $EspF_U$ to the bacterial attachment site, cytosolic side, through the binding of a PxxP motif in $EspF_U$ to the IRSp53 SH3 domain. Through the use of the same helical SLiM present in NCK (ELM:LIG_GBD_CHELIX_1), $EspF_U$ acts as a potent WASP activator, inducing the actin polymerization that contributes to the pedestal formation characteristic of EHEC infections (Cheng et al., 2008; Sallee et al., 2008). The NPY SLiM, although not yet experimentally validated in any human protein, is potentially functional in proteins like SHANK2 or the microtubule-binding CLIP-associating protein 1 (CLASP1), based on protein conservation and functional association (de Groot et al., 2011). The putative NPY motif in ACE2 is conserved in all analysed mammalian and bird homologs (Figure 3), suggesting a direct interaction with host I-BAR-containing proteins such as IRSp53 or IRTKS, which are expressed in lung tissues (Uhlén et al., 2015).

The I-BAR domain-binding motif in the cytosolic region of ACE2 could be relevant for SARS-CoV-2 infection in the following scenario. During viral cell entry, the NPY motif could recruit I-BAR-containing proteins such as IRSp53 or IRTKS, resulting in membrane protrusion formation that could be exploited for viral entry or in cell to cell transmission. It is known that the hijack of the filopodia formation network is beneficial for the entry and spreading of many enveloped viruses (reviewed in Chang et al., 2016), but whether this process is active during coronavirus infection is still unclear. A second route might cooperate with the NPY motif in the recruitment of actin cytoskeleton components. A direct interaction between the SARS-CoV Spike protein cytosolic side C-terminal domain and the Ezrin FERM domain can occur during the opening of the viral fusion pore and has been proposed to restrain viral infection (Millet et al., 2012). Ezrin is a protein involved in cell morphology and apical membrane remodelling that acts as a membrane-cytoskeleton linker. Ezrin recruits F-actin through its C-terminal domain, and can also bind to IRSp53 located at negatively curved membranes (Saleh et al., 2009; Tsai et al., 2018), suggesting that while the NPY motif acts at earlier stages of viral attachment, the Spike protein/Ezrin interaction might work during or after viral fusion, to promote the recruitment of actin regulatory components to viral fusion sites.

### *The apoPTB domain-binding motif*

Certain members of the PTB domain family were discovered to bind to phosphorylated NPxY motifs, hence the designation Phospho-Tyrosine Binding domain (Zhou et al., 1995). The NPxY motifs in cytosolic tails of receptors, including integrins, are regarded as endocytosis sorting signals (Bonifacino and Traub, 2003). It was later discovered that PTB domains in the internalisation adapter protein Dab1 could also bind non-phosphorylated Nxx[FY] motifs (apoPTB motif) and that this might be the case for the majority of PTBs (Uhlik et al., 2005). Representative receptors with apoPTB motifs are in the database entry ELM:LIG_PTB_Apo_2. In ACE2, the core NxxF motif is conserved in all land vertebrates (Figure 3). For Dab1 apoPTB motifs, there is a hydrophobic requirement two residues before the Asn. In ACE2, this is predominantly a charged residue: Therefore, if this strikingly conserved NxxF is an apoPTB motif, it should bind a protein other than Dab1 and proteins with related specificities. The apoPTB motif binds as a short β-strand (β-augmentation) followed by a β-turn. Proline is rejected at one strand-forming position and therefore the regular expression for this motif would be [^P].N..F for land vertebrates or [EQ].N..F for mammals. As with the phosphorylated versions, the apo-motifs are tightly connected to endocytosis (Uhlik et al., 2005).

### *The LC3-interacting region (LIR) autophagy motif*

Autophagy, the recycling of cellular material, is vital for cellular homeostasis. Many pathogens must control the autophagy response to establish productive infection (Deretic and Levine, 2009). It has been shown that coronaviruses, including human CoVs, subvert autophagy components to promote viral replication at DMVs associated to the RTC (Cottam et al., 2014; Fung and Liu, 2019; Gassen et al., 2019; Reggiori et al., 2010). The LIR motif is required for the interaction of a target protein with Atg8 or its homologues LC3 and GABARAP to facilitate autophagy of the target via the autophagosome (Birgisdottir et al., 2013). The LIR motif has been catalogued in the ELM resource entry ELM:LIG_LIR_Gen_1 which detected a candidate motif in the human ACE2 cytosolic tail sequence (Figure 3). After the LIR motif was annotated in ELM, a more recently solved LC3-LIR structure (PDB:5cx3) showed that the interacting peptide is longer, with one or two more hydrophobic interactions (Olsvik et al., 2015). LIR enters a hydrophobic groove bordered by positively charged residues. A core [WFY]xx[ILMV] enters the deepest part of the groove. On either side of the core, the interacting residues can be flexibly spaced. The core must be preceded by a negatively-charged residue (which might be enabled by phosphorylation). Further, the motif core is followed by a flexibly spaced hydrophobic residue. There is often a negatively-charged residue preceding this hydrophobic position: it can make favourable interactions with counter charges but is not an absolute requirement, so is not included in the revised motif pattern. Based on the structure (PDB:5cx3) and some SPOT arrays (Alemu et al., 2012; Olsvik et al., 2015; Rasmussen et al., 2019), the regular expression [EDST].{0,2}[WFY][^RKP][^PG][ILMV].{0,4}[LIVFM] matches the motifs annotated in ELM. The revised motif is conserved in the mammalian ACE2 cytosolic tail, but not in birds or reptiles. The ACE2 LIR motif candidate can potentially enable the incoming coronavirus to attract autophagy elements such as LC3 to the structures where the virus replicates and assembles. In line with this, a non-lipidated form of the LC3 protein has been shown to be associated with the RTCs of MHV and SARS-CoV (Cong et al., 2017; Prentice et al., 2004; Reggiori et al., 2010). This brings up the interesting possibility that ACE2 remains associated with the membranous structures where SARS-CoV-2 replicates at later infection stages, assisting in the repurposing of autophagy components required for viral replication.

### *The TxF$ PDZ-binding motif*

Amongst other motif-binding modules, PDZ domains come in great abundance in human and other multicellular animals (Ernst et al., 2009). PDZ domains take part in a variety of biological processes including cellular signalling and activity at the synapse (Manjunath et al., 2018). These domains bind SLiMs by β-strand augmentation which are called PDZ-binding motifs or PBMs, most commonly known to be found in the C-terminus of fully or partially disordered proteins. These interactions are widely studied and their link to various diseases and infections has been previously established (Christensen et al., 2019). A PBM candidate is also found in the very C-terminus of the cytosolic tail of vertebrate ACE2 proteins (Figure 3). Motifs following a pattern [ST].[ACVILF]$ are a common PDZ-binding motif variant, described in the ELM resource entry ELM:LIG_PDZ_Class_1. There are multiple functional examples of this motif. However, in the ACE2 protein the matching sequence is not characterized. ACE2 has a disordered tail facing the cytosol, where a number of different PDZ domains could be its potential binders (Manjunath et al., 2018).

Two PDZs in two different adapter proteins – Na(+)/H(+) exchange regulatory cofactor NHERF3 and SH3 and multiple ankyrin repeat domains protein 1 SHANK1 – have been previously identified to be able to bind TxF$ sequences (“$” stands for the C-terminal end) (Ernst et al., 2014), which makes them both candidates for an interaction with the ACE2 C-terminus. NHERF3 is co-localised with ACE2 in intestinal tissue, and its PDZ domains were previously validated to interact with PBMs in transmembrane proteins on the cytosolic side of the membrane (Gisler et al., 2003), so it is possible they come in proximity with the ACE2 tail containing the TxF$ motif, and possibly bind it as a part of ion exchange regulation of small molecule transport activities. NHERF3 is known for its involvement in sodium ion-dependent transporter activity (Srivastava et al., 2019), and ACE2 was also shown to interact with a sodium-dependent transporter (Yan et al., 2020), which could be one of the leads towards unravelling the possible interaction between

NHERF3 and ACE2. NHERF3 (gene name PDZK1) and ACE2 share a network of experimentally validated protein-protein interactions localized at the cell membrane (Figure 4). These proteins share localisation and function characteristics, and some of them could turn out to be the missing puzzle pieces to connect ACE2 and NHERF3 as possible interactors.

All in all, whether NHERF3 is the PDZ-containing protein interacting with ACE2 or not is an open question, but there should be very little doubt that ACE2 exhibits PDZ-binding activity in the cell, since the motif is highly conserved, very specific and already known to appear in membrane-associated proteins with C-terminal tails facing the cytosol.

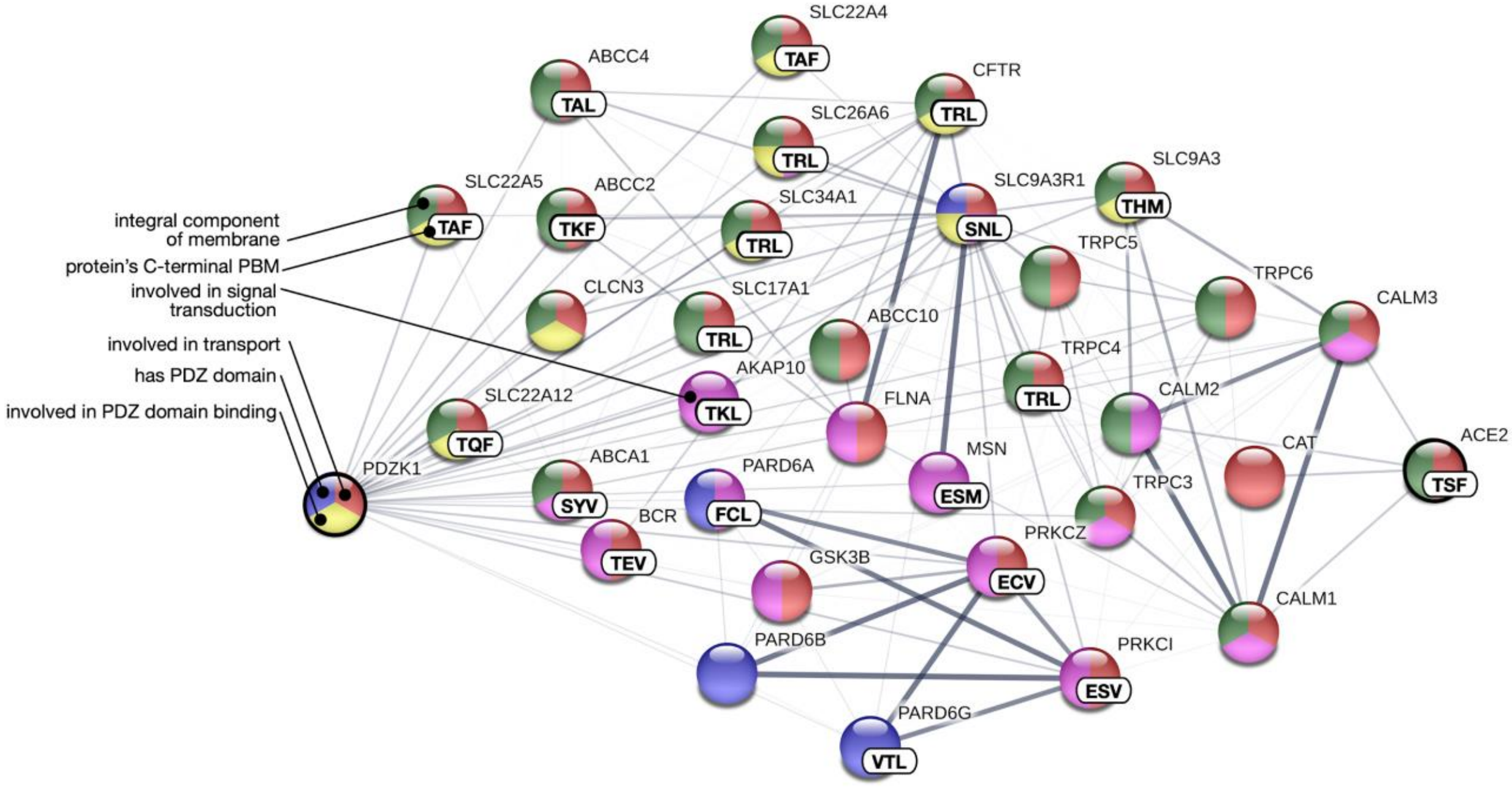

***Figure 4: Experimentally validated interaction network between NHERF3 (PDZK1) and ACE2.*** *C-terminal PBMs are displayed in white boxes. Common functionalities of the proteins marked by different colours. Thickness of nodes is proportional to the confidence behind the experimental evidence. The network was built using the STRING resource (https://string-db.org) (Szklarczyk et al., 2019).*

### *The phosphorylated Tyr781 in the ACE2 tail*

Tyrosine 781 is a part of the motif patterns for four of the motifs listed above but must be phosphorylated to act as an SH2-binding motif. Therefore, we searched the ACE2 literature for reports of phosphorylation but were unable to find any with strong site identification. Examination of the human ACE2 entry in the database PhosphoSitePlus (Hornbeck et al., 2019) revealed that high-throughput (HTP) phosphoproteomic studies, but no low-throughput (LTP) studies, identify pTyr781. As shown in Figure 5, thirteen HTP measurements identified phosphorylation at Tyr781 and this residue is the only ACE2 phosphosite that is highly reproducible across multiple HTP datasets. For example, pTyr781 was one of 318 unique phosphopeptides belonging to 215 proteins analysed from an erlotinib-treated breast cancer cell line model (Tzouros et al., 2013). Therefore, this site indeed fulfils the phosphorylation requirement to be an SH2-binding motif.

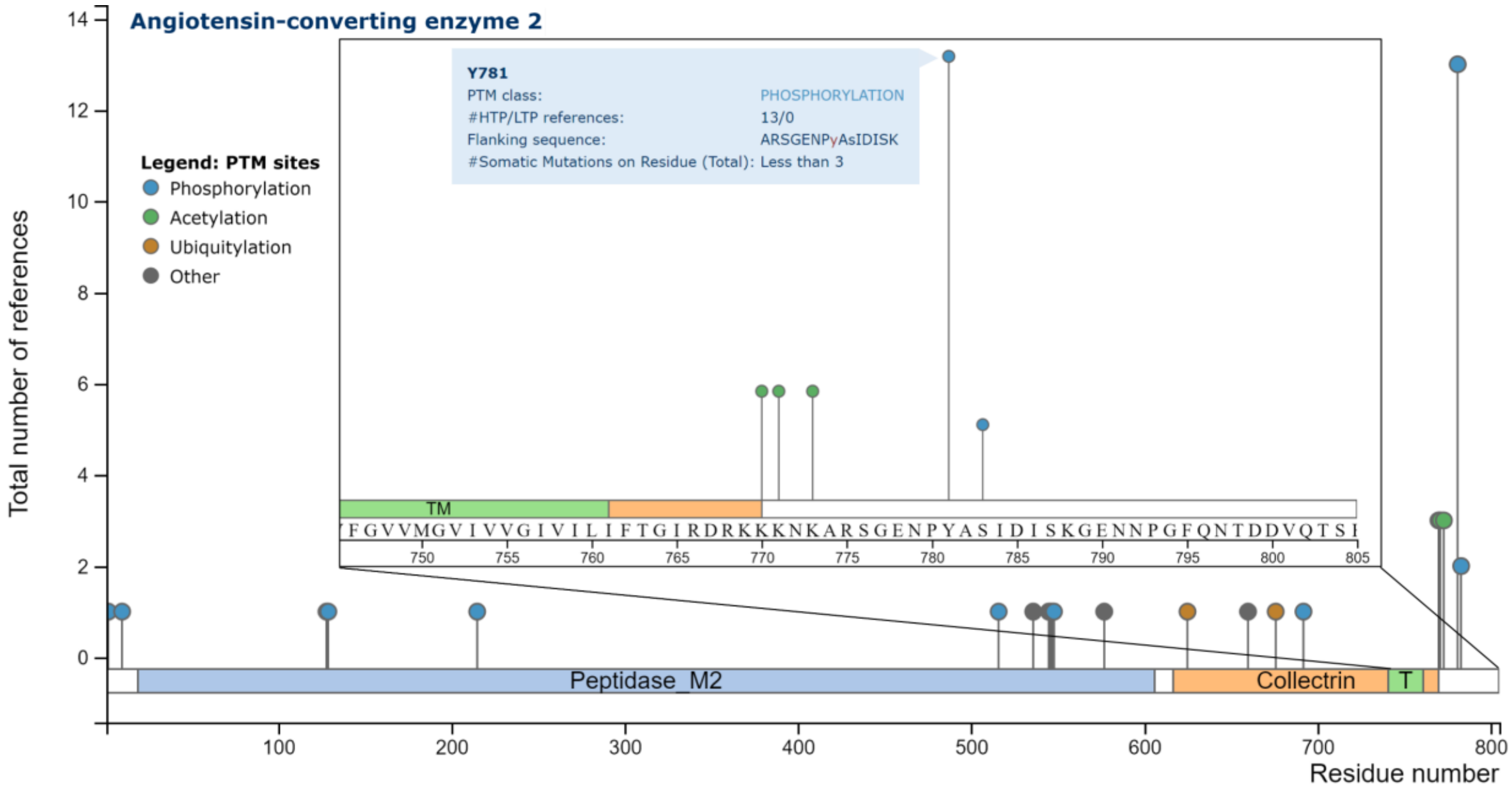

***Figure 5: The summary for the ACE2 C-terminal tail provided by PhosphoSitePlus.*** *No low-throughput (LTP) studies have been recorded in the database for ACE2. Thirteen high-throughput (HTP) studies have identified pTyr781. Phosphosites reported in the extracellular part of ACE2 have only been reported once each and therefore are likely to be misidentified peptides.*

### *A potential four way molecular switch at Tyr781 in the ACE2 tail*

As described above, four sequence motifs overlap in the region surrounding Tyr781: the YxxPhi endocytic sorting signal (ELM:TRG_ENDOCYTIC_2), an NCK SH2 binding motif (ELM:LIG_SH2_NCK_1), an NPY I-BAR binding motif (ELM:LIG_IBAR_NPY_1) and the LIR autophagy motif (ELM:LIG_LIR_Gen_1). While the YxxPhi, NPY and LIR motifs require a non-phosphorylated state of Tyr781, the NCK motif requires Tyr781 phosphorylation, creating the opportunity for a four way molecular phospho-switch acting in this region of ACE2 that directs different steps of the SARS-CoV-2 infection cycle. The state of this switch can be controlled by tyrosine kinase activity involving the Src/Abl and other tyrosine kinases known to be upregulated during endosomal processes (Reinecke and Caplan, 2014) and viral infection (Davey et al., 2011) including in coronaviruses (Coleman et al., 2016; Dong et al., 2020; Shin et al., 2018; Sisk et al., 2018). Similar two-way switches have been described before, as with the CTLA-4 receptors, where Src-family tyrosine kinases dictate the binding preferences of overlapping YxxPhi and SH2 binding motifs. In the non-phosphorylated state endocytosis is favoured, whereas T cell activation brings about Tyr phosphorylation, shutting down endosomal recycling and initiating signalling through the recruitment of SH2-domain containing proteins (Bradshaw et al., 1997; Miyatake et al., 1998; Ohno et al., 1996; Owen and Evans, 1998; Shiratori et al., 1997).

During early stages of viral infection, and following viral attachment to the host cell membrane, the YxxPhi motif could activate the early events of receptor-mediated endocytosis by binding the μ subunit of the AP2 complex, which in turn recruits clathrin and other endocytic components to the viral attachment sites. Following the formation of the clathrin coat and initial invagination, actin polymerization is required to assist in the internalization of the endocytic vesicles. In addition, some viruses can 'surf' along filopodia by myosin-mediated actin cytoskeleton movements that transport the viral particles to the entry sites at the cell body, ultimately increasing their entry rate (reviewed in Chang et al., 2016). The actin hijack related to endocytic uptake and the formation of membrane protrusions could be coordinated in one or several stages by the NCK SH2 and NPY motifs respectively, by recruiting WASP and I-BAR proteins to the viral attachment site. This might be enacted by a Tyr781 phosphorylation switch that is regulated in time with early attachment characterized by non-phosphorylated Tyr781 that allows the YxxPhi and NPY motifs to be active, and a later phase where Tyr781 becomes phosphorylated, inactivating the YxxPhi and NPY motifs and bringing the NCK SH2 motif into play. An alternative scenario might be enabled by the multimeric nature of the Spike protein and by attachment of several viral particles to a membrane domain, leading to adjacent ACE2 tails on the intracellular side that expose both phosphorylated and non-phosphorylated motifs, allowing these three signalling steps to take place simultaneously. The presence of several parallel routes for the recruitment of cytoskeleton components involving the NPY and NCK motifs could provide robustness needed to ensure the actin-reorganization required for the uptake of virus-containing vesicles into the cytosol. Following endocytosis and fusion, viral components are released into the cell and viral replication takes place. During this phase, the last component of the switch could come into play, when the ACE2 protein that remains bound to Spike protein-coated membranes could promote the hijack of autophagy components necessary to assemble the viral replication factories. At this time, non-phosphorylated ACE2 tails would recruit LC3 to replication structures through their LIR motifs.

## Known and candidate motifs in the integrin β tails

Integrin β tails are short cytosolic C-terminal intrinsically disordered regions, similar to the analysed region of ACE2. The two most probable integrin β subunit candidates at play in SARS-CoV-2 viral entry are β3 and β1. The C-terminal tail of both subunits share a high degree of sequence similarity, and similarly to ACE2, contain several known and candidate SLiMs (see Table 1 and Figure 6) that propagate signals in the cytoplasm and regulate integrin activity not just through intracellular pathways, but also changing the structural state of the ectodomains determining ligand binding capacity (Anthis and Campbell, 2011).

### *The membrane proximal tyrosine kinase interacting region*

Integrin β tails contain a highly charged patch in their membrane proximal region. This region is indispensable for the interaction between integrins and tyrosine kinases, including the Src-family kinase Fyn (Reddy et al., 2008) and the focal adhesion kinase (FAK), most probably via the direct interaction with paxillin (Liu et al., 2002). Through these interactions, integrins regulate cytoskeletal remodelling (Lv et al., 2016) and the promotion of cell survival (Tang et al., 2007), as well as regulation of focal adhesion assembly and cell protrusion formation (Hu et al., 2014). In turn, FAK regulates integrin recycling and endosomal trafficking (Mana et al., 2020; Nader et al., 2016).

Currently, there is no consensus sequence motif describing these interactions, although a definition of HDR[KR]E has been proposed (Legate and Fässler, 2009), fitting integrins β1, β3, β5 and β6. This motif is under heavy regulation by several mechanisms. First, the interaction with tyrosine kinases seem to involve additional residues N-terminal of the charged motif core – most notably the conserved lysine preceding the hydrophobic patch (Reddy et al., 1998) – that are only accessible in the active state of the integrin dimer, as these regions are buried in the membrane otherwise (Stefansson et al., 2004). Second, the D residue of the motif forms a salt bridge with the cytosolic tail of the α subunit of the integrin in the inactive conformation of the receptor. Thus, this motif region is dependent on integrin activation regulated by ligand binding and intracellular interactions mediated by the C-terminally located NPxY motifs.

### *The NPxY motifs*

Both the β1 and β3 tails contain two regions that match the apoPTB motif defined in the ELM resource (ELM:LIG_PTB_Apo_2). Furthermore, these regions are known to have Tyr phosphorylation, matching the phosphorylated motif definition as well (ELM:LIG_PTB_Phospho_1). These regions are known to be able to form β-turns, and are recognition sites for phosphotyrosine-binding domains. NPxY motifs (or NxxY in the case of the second motif of the β3 tail) are the major sorting signals mediating interactions with FERM domains for regulating endosomal trafficking (Ghai et al., 2013). In integrin β tails these motifs recruit adaptor proteins and clathrin, serving as sorting signals (Ohno et al., 1995), and the NPxY motifs in the β1 tail have a direct connection to viral entry for reovirus (Maginnis et al., 2008).

### *Interactions mediated by the NPxY motif switches*

The first NPxY motif binds talin-1, serving as a connection between the plasma membrane and the major cytoskeletal structures (Horwitz et al., 1986). Considering the expression profiles of talins, the most likely interaction partner of lung-expressed integrins is talin-1. Talin-1 contains a FERM domain, similarly to Ezrin, which establishes a direct interaction with the SARS-CoV Spike protein upon viral fusion (Millet et al., 2012). However, the interaction between the RBD and integrins offers the virus an earlier point of interference with the cytoskeletal system, being able to modulate it cooperatively with the ACE2 actin-regulatory elements (NPY and NCK SH2 motifs) before and during cellular entry. The talin/integrin interaction however presents a feedback loop: the binding of talin on the cytoplasmic side induces a structural rearrangement on the ectodomains of integrins, enabling a higher affinity interaction with RGD motif containing ligands (Tadokoro et al., 2003).

The first NPxY motif is also a binding site for Dok1, a negative regulator of integrin activation. Dok1 is in direct competition with talin for binding integrins (Tadokoro et al., 2003). The competition is fundamentally influenced by phosphorylation on Tyr773 of the NPxY motif. The non-phosphorylated motif has a higher affinity towards talin, whereas phosphorylation prefers Dok1 (Oxley et al., 2008); thus Tyr773 acts as a phospho-switch that regulates integrin activation.

The first NPxY motif also presents a site for a largely phosphorylation-independent interaction with the Integrin Cytoplasmic domain-Associated Protein-1 (ICAP-1). ICAP-1 is a fundamental regulator of the assembly of focal adhesions (FA) and ICAP-1 knockdown reduced FA assembly (Alvarez et al., 2008), possibly working in conjunction with the membrane-proximal charged motif. ICAP-1 seems to be specific for β1, and hence the therapeutic considerations for targeting this pathway requires the verification of the type of integrins expressed on AT2 cells (and other related cell types).

The second NPxY motif is a binding site for kindlin (Ma et al., 2008). This interaction requires the integrin tail to be non-phosphorylated and phosphorylation on Tyr785 (for β3) can switch off the interaction with kindlin-2 (Bledzka et al., 2010). Kindlin binding (together with talin binding) is a crucial step in integrin activation, and hence regulates the availability of integrins for extracellular ligands (Herz et al., 2006), and was also suggested to play a role in TGF-β1 signalling (Kloeker et al., 2004).

### ***Building a complex switching mechanism from two phospho-switches***

The two NPxY(-like) motifs in the integrin β tails not only constitute two separate phospho-switches, but also act in synergy to give rise to more complex regulation. Filamin and the PTB domain region of Shc1 each bind to both NPxY motifs (Deshmukh et al., 2010; Liu et al., 2015). Shc is an adaptor protein playing a key role in MAPK and Ras signalling pathways, and its interaction with integrin β3 requires both phosphorylations on Tyr773 and Tyr785 (Audero et al., 2004; Deshmukh et al., 2010). In contrast, binding of the Ig domain of filamin-A requires both tyrosines to be in a non-phosphorylated state. The filamin-A interaction can be considered as a main shutdown switch in integrin signalling, as this interaction induces the closed conformation of the integrin ectodomains, decreasing the chance of ligand binding (Liu et al., 2015). In addition, binding partners utilizing both NPxY motifs may also serve as stronger modulators of endosomal trafficking, switching on enhanced signals.

### *Potential LIR autophagy motif in integrin β3*

The connection between autophagy and cell adhesion has already been described, showing that both reduced FAK signalling (Sandilands et al., 2011) and detachment from the extracellular matrix via integrins (Vlahakis and Debnath, 2017) enhances autophagy. Atg deficient cells have enhanced migration properties, and at the molecular level there seems to be a direct connection between Atg proteins and integrins as well: autophagy stimulation increases the co-localization of β1 integrin-containing vesicles with LC3-stained autophagic vacuoles, while autophagy inhibition decreases the degradation of internalized β1 integrins (Tuloup-Minguez et al., 2013). In Drosophila cells it has been shown that the Wiskott-Aldrich syndrome protein and SCAR homolog (WASH) plays a connecting role between integrin recycling and the efficiency of phagocytic and autophagic clearance (Nagel et al., 2017). However, molecular details about how this connection is brought about are unclear.

Sequence analysis of integrin β3 tails show a potential Atg-interacting LIR motif, similarly to the ACE2 tail. Low throughput phosphorylation assays have determined that both Tyr773 and Tyr785 (required for the functional PTB NPxY motif) are in fact phosphorylated in vivo. However, such assays have also determined additional phosphorylation sites in the β3 tail, Thr777, Ser778, Thr779 and Thr784. These phosphorylations aren't connected to the NPxY motif switches in any known way. However, the three phosphorylations between residues 777-779 and the following sequence region matches the LIR autophagy motif also found in the ACE2 tail. While the current motif definition does not exactly fit the β1 tail, there is also low throughput phosphorylation assay data (Wennerberg et al., 1998) for the existence of these phosphorylations in the corresponding residues, hinting at the possibility of the presence of a slightly modified motif. For both β1 and β3 tails, the phosphorylation provides the negative charge required N-terminal of the FxxIxY LIR motif hydrophobic core. Phosphopeptides spanning the candidate region should reveal whether the LIR motif-like region is a functional Atg-binding site in integrin β1, and also whether the multiple phosphorylations act as a rheostat, modulating the affinity of the interaction. The motif found in integrin β3 is also present in integrin β2, and the motif candidate identified in integrin β1 is also present in integrin β6.

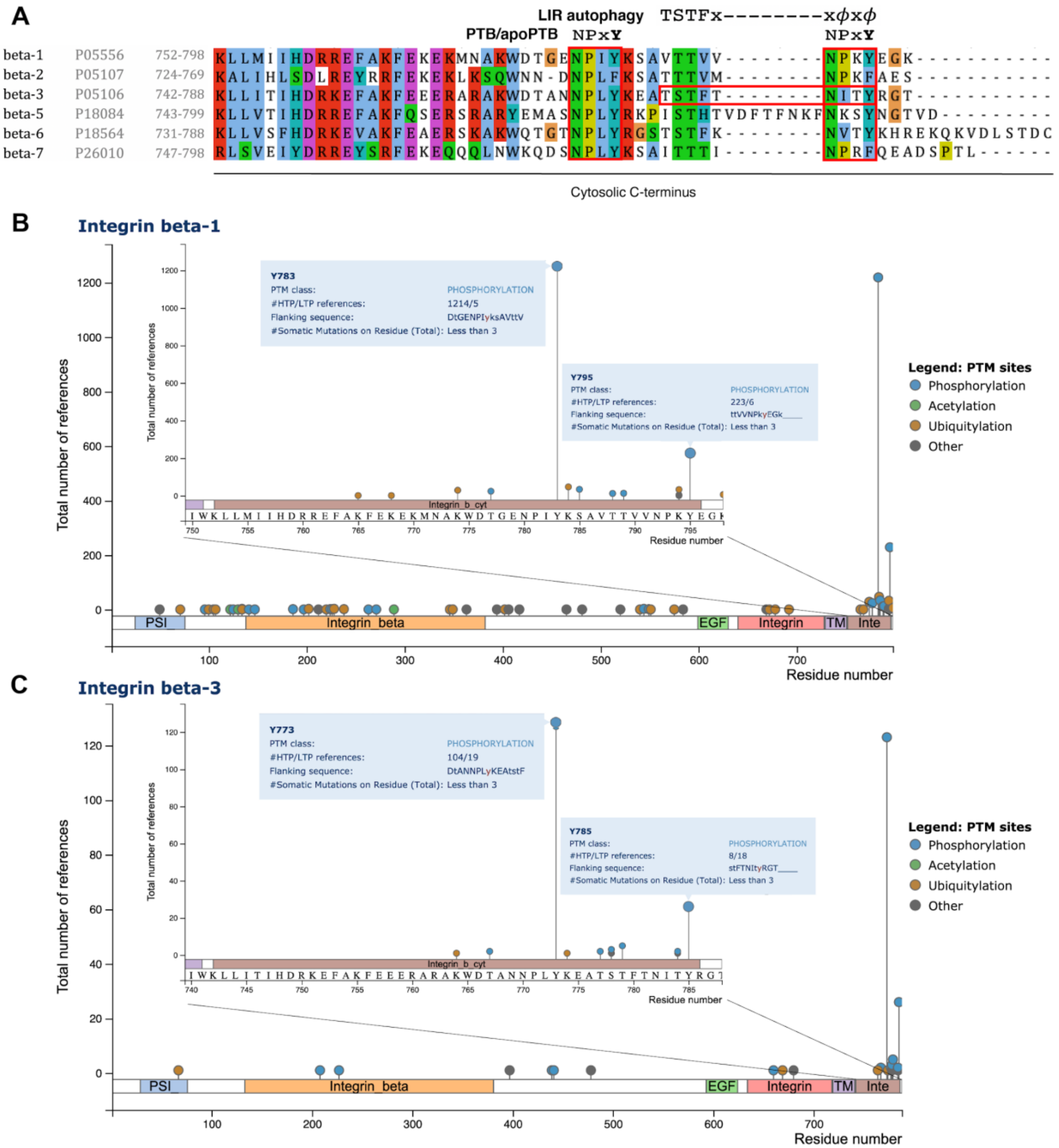

***Figure 6: Alignment of human integrins illustrating conserved motifs in the cytosolic C-terminal tail**. A) Multiple sequence alignment of human integrin C-terminus regions, not including the two divergent β tails (β4 and β8). The alignment shows their motif conservation of the NPxY and LIR motifs (key residues displayed above). Red boxes mark the conservation range of the PTB motif in all sequences and the location of the LIR motif in integrin β3. Protein names, UniProt IDs and sequence numberings are listed on the left side of the alignment. B-C) Summary of the PTMs on the integrin β1 and β3 C-terminal tails. Details of the experimental evidence for the PTB tyrosine phosphorylations are highlighted: pY783 & pY795 for β1 and pY773 & pY785 for β3. Graphs obtained from PhosphoSitePlus (09/04/2020).*

## Potential synergy between the ACE2 and integrin intracellular motifs

Bringing together the candidate SLiMs identified in the integrin β and ACE2 tails potentially strengthens the functional links between them, and provides an emergent picture of SLiM-driven cooperative switches driving viral attachment, entry and replication (Figure 7). Following attachment of Spike to the receptors, the two NPxY motifs in the integrin β subunit could act cooperatively with the apoPTB and YxxPhi motifs in ACE2 as sorting signals that mediate the internalization of viral particles into endosomes. The presence of several endocytic motifs in close proximity would strengthen the interaction with the endocytosis apparatus creating a high-avidity environment for recruitment of RME components (Bonifacino and Traub, 2003). During this time, the phosphorylated integrin NPxY motifs would also reinforce viral attachment through inside out signalling, stabilizing the integrin ectodomain in the open, high affinity ligand-binding conformation. As discussed previously, RME also involves the recruitment of adaptor molecules that activate rearrangements of the actin cytoskeleton required for the internalization of the endocytic vesicle. At this stage, the NPY and NCK SH2 motifs in ACE2 would recruit several molecules that mediate actin polymerization signalling, prominently I-BAR containing proteins IRSp53 and IRTKS, and the WASP-Arp2/3 complex. While most of this actin signalling would serve to allow viral entry, additional actin recruitment processes could occur following viral fusion, such as that initiated by the interaction between the Spike protein and Ezrin. Finally, at later stages of infection, both integrins and ACE2 might remain attached to virus-associated DMVs and other replication-competent membranes where the RTC assembles. At this stage, ACE2 and integrins might cooperatively mediate the recruitment of autophagy components such as LC3, through the LIR motifs located in the cytosolic tails of both molecules.

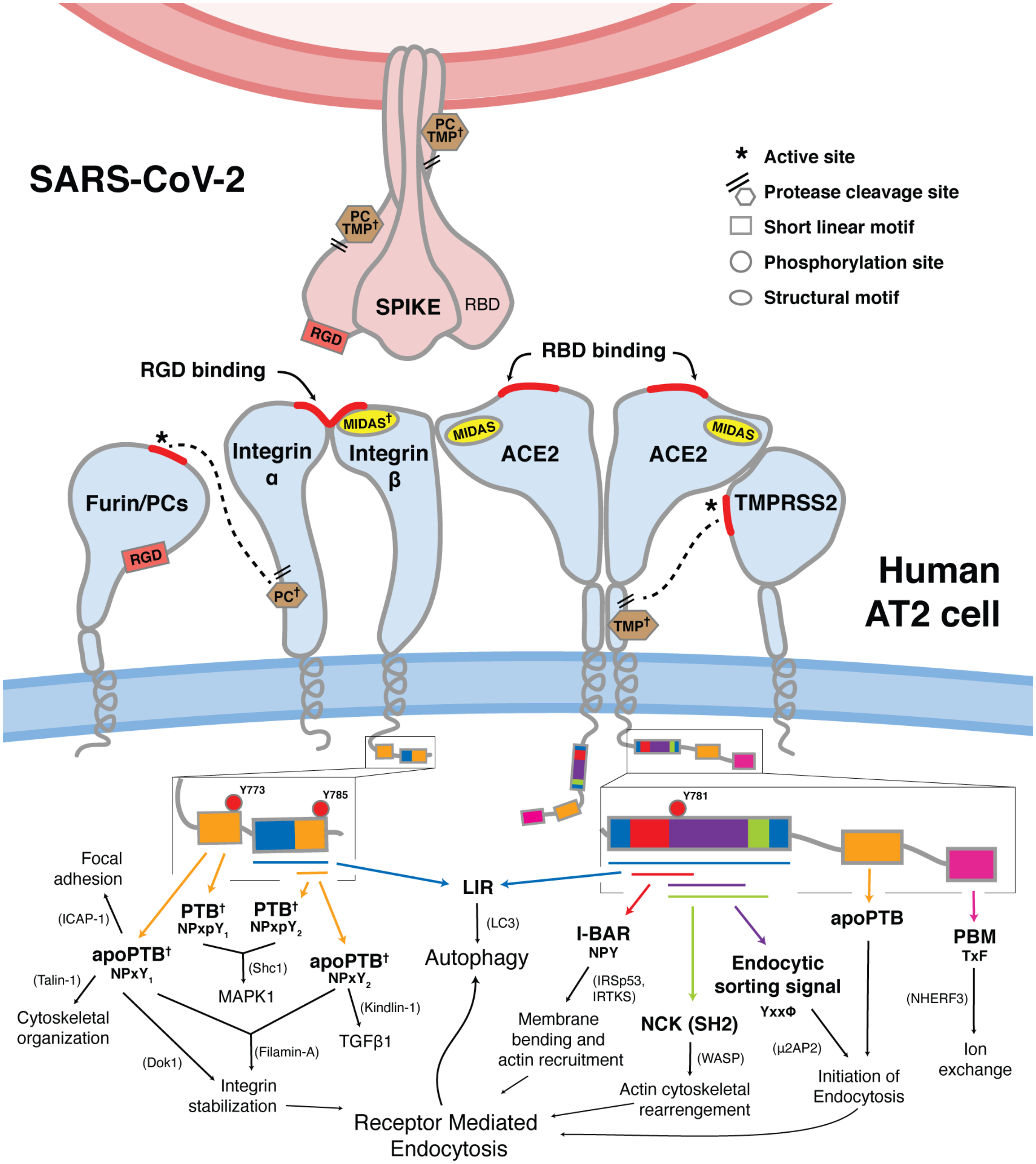

***Figure 7: Model of the proposed interplay between motifs in the interface between SARS-CoV-2 and human AT2 cell to achieve receptor mediated endocytosis****. Receptors of the SARS-CoV-2 (pink) and human AT2 cell (light blue), motifs involved in viral recognition and entry are shown in boxes. Elements shown in one of the monomers of a homotrimer (Spike) or homodimer (ACE2) are also present in the other proteins forming that complex. Lines below motif boxes represent each of the overlapping motifs in that specific region. Arrows indicate the related cellular process and the protein known to interact with their respective motif is indicated in*

*parenthesis. Phosphorylation sites are shown as red circles with the respective sequence position indicated. For the integrin β tail, the PTB/apoPTB phospho switch is depicted as two separate versions of the same motif region, and the subscripts represent the motif order in the sequence. The colour code is as follows: cleavage sites (brown), for motifs: apoPTB/PTB (orange), endocytic sorting signal (purple), I-BAR binding (red), LIR (blue), MIDAS (yellow), NCK (green), PBM (magenta), RGD (light red). SLiMs mediating interactions are marked with coloured aquares, protease cleavage sites are marked with hexagons, structural motifs are marked with ovals. Motifs marked with † are experimentally validated.*

## Short linear motifs and their potential therapeutic implications

The analysis of candidate SLiMs in ACE2 and integrins suggests that SARS-CoV-2 hijacks both receptors, co-opting their SLiMs to drive viral attachment, entry and replication. This creates an opportunity for drugging these interactions, or the processes they control through Host Directed Therapies (HDTs), to prevent viral entry. A list of potentially useful drugs is presented in Table 2, together with ChEMBL accessions (Mendez et al., 2019).

The sequence RGD is used by a large number of viruses for cell attachment, via integrins (Hussein et al., 2015). RGD mimics have been developed as inhibitors of integrin-extracellular matrix protein interaction for a variety of diseases. A cyclic RGD peptide (c-RGDfV, cilengitide) has been developed clinically for glioblastoma treatment and other cancers. It proved safe, but did not enhance the survival benefit (Stupp et al., 2014). SARS-CoV-2 has a unique RGD sequence in the ACE2 binding region of its Spike protein. It has been speculated that integrins may have a potential role for infectivity (Sigrist et al., 2020). Therefore, integrin inhibitors like RGD mimetics might be able to block the RGD binding site(s) on target cells and block the attachment of the virus. Another application that has been suggested is bacterial sepsis (sepsis is also a dreaded complication in COVID-19 patients), and experimental evidence in animals is available (Garciarena et al., 2017). Cilengitide is relatively specific for integrin αvβ3 and also active on αvβ5, αvβ6. The antibody abituzumab (aka DI 17E6) is a pan-av antibody, i.e. also active against other αv integrins, and may consequently be better suited for blocking virus entry. It has been clinically tested in several cancer indications (Élez et al., 2015; Hussain et al., 2016). Cilengitide and abituzumab are made available for in vitro testing in the context of the SARS-CoV-2 pandemic by the company who developed it initially (contact: compound_donation@merckgroup.com).

As discussed above, tyrosine kinase mediated phosphorylation plays an important role in virus entry and maturation, and several tyrosine kinase inhibitors have been developed in the clinic and show effects on viral infection. For example, saracatinib, a Src and Abl inhibitor which has completed several clinical trials, including some cancers, inhibited replication of different coronaviruses including MERS-CoV, SARS-CoV and HCoV-229E in cell culture infection experiments (Shin et al., 2018). After internalization and endosomal trafficking, imatinib, an Abl inhibitor, prevented fusion of SARS-CoV and MERS-CoV virions at the endosomal membrane in infected cell culture experiments (Coleman et al., 2016). Using the avian model virus, IBV, imatinib and two other Abl inhibitors, GNF2 and GNF5, prevented the fusion of the Spike protein to the membrane of the target cell as well as cell-cell fusion and syncytia formation (Sisk et al., 2018). More recently, tyrphostin A9, a platelet-derived growth factor receptor (PDGFR) tyrosine kinase inhibitor came out from a high-throughput screening using cytopathic effect as readout, it also showed *in vitro* inhibitory capacity to transmissible gastroenteritis virus (TGEV), an alpha coronavirus that infects pigs (Dong et al., 2020). The authors also showed that tyrphostin A9 has a broad spectrum, being active against three other tested coronaviruses: MHV in L929 cells, porcine epidemic diarrhea virus in Vero cells and feline infectious peritonitis virus in CCL-94 cells. The mode of action was found to be through p38 MAPK, at the post-adsorption stage. As FAK has been implicated in viral entry for other viruses including influenza A (Elbahesh et al., 2014), experimental drugs targeting FAK, including some in clinical trials (de Jonge et al., 2019), can be considered for studying the potential Spike-induced integrin signalling. Currently, 39 tyrosine kinase inhibitors are approved by the FDA: eleven target non-receptor protein-tyrosine kinases and 28 inhibit receptor protein-tyrosine kinases (Roskoski, 2020). Consequently, tyrosine kinase inhibitors may be good candidates to test for their effect on SARS-CoV-2.

A number of protease inhibitors are currently discussed for SARS-CoV-2 treatment. Serine protease inhibitor camostat mesylate is active against TMPRSS2 and blocks cell entry (Hoffmann et al., 2020). Only the Spike protein of SARS-CoV-2 contains a furin cleavage sequence (PRRARS|V). Consequently, furin convertase inhibitors are considered as antiviral agents (Shiryaev et al., 2007).

Many viruses enter the cell via endocytosis, and a number of candidate SLiMs relevant for SARS-CoV-2 infection are related to endocytosis (see above). Chlorpromazine, an antipsychotic (via dopamine D2 antagonism) dating from the 1950s, is also a potent endocytosis inhibitor (which may explain some of its marked side effects, which can include low white blood cell levels). It promotes the assembly of adaptor proteins and clathrin on endosomal membranes thus depleting them from the plasma membrane, leading to a block in clathrin-mediated endocytosis (Wang et al., 1993). The potential use of endocytosis inhibitors such as Amiodarone (Stadler et al., 2008) and chlorpromazine in coronavirus infection is further discussed here (Yang and Shen, 2020).

The situation with targeting autophagy seems unclear. Autophagy activators might help the cell to consume incoming virus - or speed up the establishment of the viral replication complexes and accelerate disease. Autophagy inhibitors might work in later stages of infection to dampen viral production, but this will depend on whether autophagy is active at the time or if the constituent components have been captured and effectively shut down. Several inhibitors/activators have been reported which can target autophagy and multiple auxiliary signals feeding into the process of autophagy (Table 2). One such axis is via the mTORC1 complex. Active mTORC1 keeps the autophagy process inhibited by phosphorylating the ULK complex which is a key regulator in autophagy. Inhibition of mTORC1 activates autophagy. Multiple FDA approved mTOR inhibitors are known and include Rapamycin and Everolimus. Rapamycin has been shown to be effective in cell culture for countering MERS-CoV infection and might assist in tackling SARS-CoV-2 infection (Kindrachuk et al., 2015) although the stage of infection might be crucial for the desired outcome. Simvastatin is another drug which is known to upregulate autophagy via the mTOR pathway (Wei et al., 2013). Simvastatin has also been reported to alleviate airway inflammation in a mouse asthma model (Gu et al., 2017). Another autophagy modulator is Niclosamide which regulates autophagy by targeting the autophagy regulator Beclin1 via SKP2 E3-ligase in MERS-CoV infection. In this case, reduced Beclin1 levels lead to blocking fusion of autophagosomes and lysosomes and hence the virus protects itself in the host (Gassen et al., 2019). Overall, inhibiting SKP2 by Niclosamide relieves Beclin1, allowing autophagosome-lysosome fusion and resumption of autophagy to reduce the MERS-CoV production. In addition, niclosamide valinomycin has been shown to target SARS-CoV in cell cultures as well (Wu et al., 2004).

## Discussion

A recent proteomic study expressed 26 tagged SARS-CoV-2 proteins individually to create a viral-human protein-protein interaction map (Gordon et al., 2020). As a result, 69 compounds, some being FDA approved, are candidates for drug repurposing. The Spike protein interacted with the GOLGA7-ZDHHC5 acyl-transferase complex, possibly for palmitoylation on Spike's cytosolic tail. Spike did not pull down ACE2 or any cell surface receptor protein, but these experiments are not an assay for viral entry, and therefore many protein interactions related to the viral entry mechanism are likely missing from these data. Therefore our observations of SLiM candidates in the viral attachment, entry and replication system reflect additional areas of the cell where drug repurposing for host-directed therapy might be explored. Although tyrosine kinase inhibitors have frequently been shown to dampen pathogen invasion and disease progression in cell culture, there has been little effort to move these findings into the clinic (Sámano-Sánchez and Gibson, 2020). Because of their widespread use in cancer, the safety profiles of tyrosine kinase inhibitors are well known and we wonder whether this might be a neglected opportunity.

Drugging the cell to cure the pathogen using HDTs is unlikely to fully remove a virus. This would also be undesirable, because the immune system must mount a defence in order to prevent viral reinfection. Rather, dampening viral load during viral invasion or replication should be the target, to give the host defences time to respond. It is well known that drugs like Tamiflu that slows influenza exit, and therefore entry into uninfected cells, can only have a strong effect when taken prophylactically or early in infection (Bassetti et al., 2019). Depending on the importance of integrins in SARS-CoV-2 lung cell entry, reducing viral entry is a possible role for cilengitide or other molecules that hamper integrin or ACE2 binding. An endocytosis inhibitor might play a similar role and is independent of receptor type. However, for any such inhibitor that passes the blood-brain barrier, effects on mood and other brain operations are an inevitable side effect: Even so, the endocytosis inhibitor chlorpromazine is a widely used drug with a well-known safety profile (Solmi et al., 2017).

Due to the presence of the cell attachment motif RGD in SARS-CoV-2, integrin inhibitors seem worthwhile to explore further. Cilengitide, a relatively selective integrin αvβ3 and αvβ6 inhibitor (a cyclic peptide that proved safe in patients, but failed to show a survival benefit in glioblastoma (Stupp et al., 2014)) might be useful in two phases: it could block virus attachment to target cells, and it has also been proposed as a potential treatment in sepsis (Garciarena et al., 2017). Sepsis was the most frequently observed complication among COVID-19 patients in Wuhan (F. Zhou et al., 2020). Another potential application of integrin inhibitors, especially integrin αvβ6, would be lung fibrosis - patients on respirators tend to develop lung fibrosis, and show increased αvβ6 levels (Horan et al., 2008). The antibody abituzumab (aka DI 17E6) is a pan-αv antibody with high potency on αvβ6, i.e. also active against several αv integrins, and may consequently be better suited for blocking virus entry, or may be suitable for lung fibrosis or sepsis protection. It has been clinically tested in several cancer indications (Élez et al., 2015; Hussain et al., 2016), and proved safe, but did not achieve a survival benefit in cancer.

Whether the enzymatic function of ACE2 has a role in SARS-CoV-2 infection is unknown. However, it would be readily testable with available ACE2 inhibitors, like Captopril, on the market since the 1970s (Enalapril would not be suited for in vitro testing, as it needs to be activated in the liver to become the active ingredient). It might be productive to test for synergy between molecules that block viral binding to ACE2 and molecules that block binding to integrins.

## Summary/Conclusion

We have presented evidence at the sequence level for SLiMs in ACE2 and β integrins with the potential to function in viral attachment, entry and replication for SARS-CoV-2. We identified several candidate molecular links and testable hypotheses that might help uncover the (still poorly understood) mechanisms of SARS-CoV-2 entry and replication. Because these motifs belong to host proteins acting as viral receptors, they are not revealed by virus-centred proteomic assays. Most of these putative motifs lack direct experimental evidence. That they may well be functional, however, is indicated by sequence conservation, in some cases for hundreds of millions of years. In addition, these motifs are in appropriate cellular contexts to interact with their respective partner proteins. Experimental validation will yield insights into receptor-mediated endocytosis for SARS-CoV-2 virus and, in addition, for the role of ACE2 in the normal cell, where it surely has much more functionality than being an angiotensin converting enzyme. Overall, the collection of candidate motifs in this system suggests that a range of host directed therapies might be explored including RGD inhibition, tyrosine kinase inhibition, endocytosis inhibition and autophagy inhibition and/or activation.

## Acknowledgements

BM has received funding from the European Union’s Horizon 2020 research and innovation programme under the Marie Skłodowska-Curie grant agreement No. 842490 (MIMIC). JČ is supported by the European Union's Horizon 2020 research and innovation programme under the Marie Skłodowska-Curie grant agreement No. 675341 (PDZnet). EM-P is a PhD student of CONICET, Argentina. RA is supported by BMBF-funded Heidelberg Center for Human Bioinformatics (HD-HuB) within the German Network for Bioinformatics Infrastructure (de.NBI #031A537B) and ELIXIR Germany. LBC is a National Research Council Investigator (CONICET, Argentina). The work was supported by Agencia Nacional de Promoción Científica y Tecnológica (PICT 2017-1924) grant to LBC. This paper is part of a project that has received funding from the European Union’s Horizon 2020 research and innovation programme under the Marie Skłodowska-Curie grant agreement No. 778247 (IDPfun) to LBC and TJG. IDPfun also funded EM-P’s placement at EMBL.

**Table 1. Known and predicted short linear motifs (SLiMs) in the SARS-CoV-2/host interaction**

| Region | Protein (UniProt accession) | Motif | ELM class[1] | Main residues[2] | Regular expression | Start | End | Sequence[3] | Binding domain[4] | Interaction partner[5] | Interaction type |
|---|---|---|---|---|---|---|---|---|---|---|---|
| Extra-cellular | SARS-CoV-2 Spike protein (P0DTC2) | RGD | LIG_RGD | RGD | RGD | 403 | 405 | RGD | Pfam:PF00362 and Pfam:PF01839 | RGD binding integrins, most probably α5β1 and αvβ3 | host:virus |
| | | Multibasic cleavage sites | - | RRxR | - | 682 | 687 | RRAR*SV | Pfam:PF00082 or InterPro:IPR001254 | Furin-like PCs / TMPRSS2 | host:virus |
| | | | - | KxxKR | - | 811 | 817 | KPSKR*SF | | | |
| | Integrin αv (similar for other α chains) (P06756) | Multibasic cleavage sites | - | xKR | - | 888 | 892 | TKR*DL | Pfam:PF00082 | Furin-like PCs | host |
| | Integrin β3 (similar for other β chains) (P05106) | MIDAS[6] | - | DxSxS | D.[TS].S | 145 | 149 | DLSYS | - | The acidic part of RGD-like ligands | host |
| | Furin (P09958) | RGD | LIG_RGD | RGD | RGD | 498 | 500 | RGD | Pfam:PF00362 and Pfam:PF01839 | Possibly RGD-binding integrin dimers | host |
| | ACE2 (Q9BYF1) | MIDAS[6] | - | DxSxS | D.[TS].S | 543 | 547 | DISNS | - | Unknown partner with acidig residue via metal ion coordination | host |
| | | Multibasic cleavage site | - | R | - | 697 | 716 | RTEVEKAIRMSRSRINDAFR | InterPro:IPR001254 | TMPRSS2 | host |
| Intra-cellular | | I-BAR binding | LIG_IBAR_NPY_1 | NPY | NPY | 779 | 781 | NPY | InterPro:IPR027681 | I-BAR domain containing proteins like IRSp53 or IRTKS | host |
| | | Endocytic sorting signal | TRG_ENDOCYTIC_2 | YⱣxΦ | Y[^P].[LMVIF] | 781 | 784 | YASI | Pfam:PF00928 | AP-2 Adapter Protein complex µ2 subunit | |
| | | NCK SH2 binding | LIG_SH2_NCK_1 | YxxΦD | (Y)[DESTNA][^GWFY][VPAI][DENQSTAGYFP] | 781 | 785 | YASID | Pfam:PF00017 | SH2 Domain of the NCK adapter protein | |
| | | LIR autophagy | LIG_LIR_Gen_1 | ExxYxxΦxΦ | [EDST].{0,2}[WFY][^RKP][^PG][ILMV].{0,4}[LIVFM] | 778 | 786 | ENPYASIDI | Pfam:PF02991 | Related proteins LC3, Atg8, GABARAP. There may be some variation in LIR motif specificity | |
| | | apoPTB | LIG_PTB_Apo_2 | Nxx[FY] | (.[^P].NP.[FY])\|(.[ILVMFY].N..[FY].) | 789 | 796 | GENNPGFQ | Pfam:PF08416 | PTB-containing protein with a preference for NxxF core motifs | |
| | | PBM | LIG_PDZ_Class_1 | TxF$ | [ST].[ACVILF]$ | 800 | 805 | DVQTSF | Pfam:PF00595 | PDZ-containing proteins with TxF$ preferences such as NHERF3 and SHANK1 | |
| | Integrin β3 (P05106) | apoPTB | LIG_PTB_Apo_2 | Nxx[FY] | (.[^P].NP.[FY])\|(.[ILVMFY].N..[FY].) | 767 | 774 | TANNPLYK | Pfam:PF00373 Pfam:PF00630 | talins (high affinity) Dok1 (low affinity) filamin-A (binding to both apoPTB motifs simultaneously) | host |
| | | | | | | 779 | 786 | TFTNITYR | Pfam:PF00373 Pfam:PF00630 | kindlin filamin-A (binding to both apoPTB motifs simultaneously) | |
| | | PTB | LIG_PTB_Phospho_1 | Nxx(Y) | (.[^P].NP.(Y))\|(.[ILVMFY].N..(Y)) | 767 | 773 | TANNPLY | Pfam:PF08416 Pfam:PF00640 Pfam:PF02174 | talins (low affinity) Dok1 (high affinity) Shc (binding to both PTB motifs simultaneously) | |
| | | | | | | 779 | 785 | TFTNITY | Pfam:PF00640 | Shc (binding to both apoPTB motifs simultaneously) | |
| | | LIR autophagy | LIG_LIR_Gen_1 | ExxYxxΦxΦ | [EDST].{0,2}[WFY][^RKP][^PG][ILMV].{0,4}[LIVFM] | 777 | 783 | TSTFTNI | Pfam:PF02991 | Atg8 protein family | host |
| | Integrin β1 (P05556) | apoPTB | LIG_PTB_Apo_2 | Nxx[FY] | (.[^P].NP.[FY])\|(.[ILVMFY].N..[FY].) | 777 | 784 | TGENPIYK | Pfam:PF00373 Pfam:PF10480 Pfam:PF00630 | talins (high affinity) Dok1 (low affinity) ICAP-1 filamin-A (binding to both apoPTB motifs simultaneously) | host |
| | | | | | | 789 | 796 | TVVNPKYE | Pfam:PF00373 Pfam:PF00630 | kindlin filamin-A (binding to both apoPTB motifs simultaneously) | |
| | | PTB | LIG_PTB_Phospho_1 | Nxx(Y) | (.[^P].NP.(Y))\|(.[ILVMFY].N..(Y)) | 777 | 783 | TGENPIY | Pfam:PF10480 Pfam:PF00640 Pfam:PF02174 | talins (low affinity) Dok1 (high affinity) ICAP-1 Shc (binding to both PTB motifs simultaneously) | |
| | | | | | | 789 | 795 | TVVNPKY | Pfam:PF00640 | Shc (binding to both PTB motifs simultaneously) | |

verified motifs (blue)
motif candidates (yellow)

[1] Motif identifier as in the Eukaryotic Linear Motif Resource
[2] x - any residue, Ᵽ - not proline, Φ - bulky hydrophobic, [FY] - F or Y, $ - C terminus, (Y) - phosphorylated tyrosine
[3] '*' marks cleavage points for protease-recognition motifs
[4] Defined using Pfam (El-Gebali et al., 2019) or InterPro (Mitchell et al., 2019), where applicable
[5] PC: proprotein convertases
[6] Not a SLiM but a structural motif

**Table 2: Drugs acting on different processes involved in viral entry and infection.**

| Name of Drug | Mode of action | Clinical status | Other details | ChEMBL ID[1] |
|---|---|---|---|---|
| **Inhibitors of viral attachment** | | | | |
| Camostat | Inhibits TMPRSS2 | Approved (Japan) | Shown to be relevant in pancreatic fibrosis | 590799 |
| Abituzumab | Integrin inhibition (**αvβ6**, pan-αv) | Phase 2 in oncology | May also be relevant in sepsis, fibrosis | 2109621 |
| Cilengitide | Integrin inhibition (**αvβ3**, αvβ5, αvβ6) | Phase 3 in oncology | May also be relevant in sepsis, fibrosis | 429876 |
| **Endocytosis inhibitors** | | | | |
| Amiodarone | Inhibits late endosomes | FDA Approved | Cell culture based evidence for SARS-CoV[2] | 633 |
| Chlorpromazine | Blocks Clathrin-mediated endocytosis | FDA Approved | Anti-psychotic drug routinely used as endocytosis inhibitor in cell culture | 823 |
| Imatinib | Abl inhibitor | FDA Approved | First line treatment for chronic myeloid leukaemia | 941 |
| Teicoplanin | Inhibits Cathepsin L (late endosomes and lysosome) | FDA Approved | Glycopeptide antibiotic | 2367892 |
| BI-853520 | FAK inhibitor | Phase 1 | FAK has been implicated in the Influenza A virus cell entry and replication[3] | 3544961 |
| Saracatinib | Src and Abl inhibitor | Phase 2 in oncology | Currently also considered for Alzheimer's disease | 217092 |
| Tyrphostin A9 | PDGF receptor kinase inhibitor (plus other activities) | Preclinical | Inhibits actin ring formation | 78150 |
| **Autophagy modulators** | | | | |
| Azithromycin | Not known | FDA Approved | Approved for multiple bacterial infections | 529 |
| Chloroquine<br>Hydoxy-Chloroquine | Lysosomal inhibition | FDA Approved | Antimalarial agent | 76<br>1690 |
| Metformin | NDUF modulation; mTOR pathway modulation (plus other activities?) | FDA Approved | Approved for type 2 diabetes | 1431 |
| Rapamycin<br>Everolimus | mTORC1 inhibition | FDA Approved | To prevent transplant rejection | 1908360<br>413 |
| Simvastatin | Autophagy upregulation via mTOR | FDA Approved | Treatment for dyslipidemia and atherosclerosis prevention | 1064 |
| Niclosamide<br>Valinomycin (VAL) | Inhibits SKP2 | Niclosamide - FDA Approved; VAL - Preclinical | Niclosamide - moderate effect; VAL targets SARS-CoV in cell culture[4] - known to inhibit SKP2 | 1448 |
| NVP-BEZ235/Dactolisib | Autophagy induction | Phase 2 | PI3k/Akt/mTOR | 1879463 |

[1]Drug details are accessible by clicking ChEMBL ids (Mendez et al., 2019); [2](Stadler et al., 2008); [3](Elbahesh et al., 2014); [4](Wu et al., 2004)